\documentclass [12pt,preprint]{aastex}
\begin{document}

\title{Merger Induced Globular Cluster Formation and Galaxy Evolution}

\author{F.D.A. Hartwick}

\affil{Department of Physics and Astronomy, \linebreak University of Victoria, 
Victoria, BC, Canada, V8W 3P6}
\begin {abstract}
Based on the earlier work of Gunn and McCrea we model the formation of globular
clusters in merging galaxies. Neutral hydrogen observations of dwarf 
irregular galaxies as well as more luminous systems are used to provide the 
key parameters of the model. The observations indicate that clusters with the 
mass of globular clusters should still be forming today. The model is 
incorporated into a phenomenological picture of galaxy evolution making use of 
a simple chemical evolution model. These results are compared to recent 
observations of the metallicity distributions of F and G stars from a recent  
large SDSS survey. The comparisons are consistent with an anisotropic 
collapse and merging of a large number of dwarf irregular galaxies for the 
formation of the Galaxy. 
\end{abstract}


\section {Introduction}

One of the remarkable observational facts about the stellar content of
galaxies is the ubiquity of globular star clusters. These clusters can be 
found in almost all but the least massive galaxies. Generally globular clusters
come in two major families of comparable mass: metal poor, old blue clusters 
and metal richer, younger, red clusters.  A comprehensive review of globular 
cluster 
systems is given by Harris (2001). Below we present quantitative arguments 
for the formation of clusters during mergers and use the observations to 
obtain the key parameters of the model. This formation picture leads naturally
to a phenomenological model for galaxy evolution. Combined with a simple 
chemical evolution model comparisons with recent data for Galactic halo stars 
are made. 

\section{Merger Induced Globular Cluster Formation}

The possibility of forming globular clusters from collisions between 
gas rich bodies was considered by Gunn (1980) and independently   
by McCrea (1982). We revive and extend the Gunn and McCrea analysis here.
\subsection{The model}
Within each of two equal mass interacting galaxies consider an undisturbed 
column of gas of
length l, cross-sectional area A and density $\rho_{o}$ meeting its counter- 
part with relative velocity 2V (see Fig.1). During the collision, the gas is 
compressed to column length $\lambda$ and density $\rho$. The shocked 
gas is assumed to cool to T$\sim10^{4}$ degrees. This assumption will 
be justified later-suffice it to say that the cooling time within the 
compressed volume ($\tau_{cool}$) must be shorter than the collision time 
($\tau_{coll}\sim l/V$) for the assumption to be valid. By conservation of mass
and for simplicity assuming A to remain constant
\begin{equation}
2\rho_{o}lA=\rho\lambda A
\end{equation}
equivalently in terms of the column density
\begin{equation}
2\Sigma_{o}=\Sigma
\end{equation}
and
\begin{equation}
\frac{\rho}{\rho_{o}}=\frac{2l}{\lambda}
\end{equation}
with the relative velocity being 2V, the relation between $\rho_{o}$ and 
$\rho$ becomes (e.g. Spitzer (1978), equation 10-24)
\begin{equation}
{\rho_{o}4V^{2}} \simeq {\rho\left(\frac{kT}{\mu m_{H}}\right)}
\end{equation}
so that
\begin{equation}
\frac{\rho}{\rho_{o}} \simeq 0.048~V^{2}/(T_{4}/\mu)
\end{equation}
where the bracketed quantity in equation (4) represents the square of the 
sound speed, and in equation (5) 
the unit of V is km/sec, T$_{4}$ is the temperature in units of $10^{4}$
degrees and $\mu$ is the mean molecular weight of the shocked gas. The 
quantity (T$_{4}/\mu$) occurs frequently in the expressions below. With 
$\mu=0.59$ a typical value for this quantity is $\sim 1.7$.
Combining (3) and (5)
\begin{equation}
\frac{l}{\lambda} \simeq 0.024~V^{2}/(T_{4}/\mu )
\end{equation}
The Jean's length, $\lambda_{J}$ is
\begin{equation}
\lambda_{J}=\left(\frac{\pi kT}{\mu m_{H}G\rho}\right)^{1/2}\simeq
\frac{246}{\rho^{1/2}}{(T_{4}/\mu )}^{1/2}~pc
\end{equation}
with $\rho$ in units of $M_{\odot}/pc^{3}$.
Equating $\lambda$ with $\lambda_{J}$ eliminating $\rho$
\begin{equation}
\lambda=\lambda_{J}=\frac{6.05\times10^{4}}{2\Sigma_{o}}(T_{4}/\mu)~pc
\end{equation}
Finally, the Jean's mass is given by
\begin{equation}
M_{J}=\rho \lambda^{3}=2\Sigma_{o}\lambda^{2}=6.05\times10^{4}\lambda(T_{4}/
\mu)~M_{\odot}
\end{equation}

If the cooling time of this gravitationally unstable mass, $\tau_{cool}$, is 
shorter than the 
free-fall time ($\tau_{ff}$) then fragmentation can occur. The cooling 
function $\Lambda(Z)$ which when 
multiplied by the square of the particle density of hydrogen gives the cooling
rate per unit volume, falls precipitously from a metallicity independent peak 
near T$\sim $T$_{4}$ to a metallicity dependent plateau below (e.g. Dalgarno
\& McCray 1972). The cooling rate from heavy elements alone at $T_{4}\sim 0.9$
is $\gtrsim 4\times10^
{-26}(Z/Z_{\odot})n_{H}^{2}$ (e.g. Fall \& Rees 1985, $\S$IV, p. 22). From 
$\tau_{cool}=
1.5nkT/(\Lambda n_{H}^{2})$ and with the density in $M_{\odot}~pc^{-3}$ the 
cooling time scale becomes 
$\sim 6.96\times10^{4}(Z/Z_{\odot})^{-1}(T_{4}/\mu)/\rho$ yrs. Similarly 
$\tau_{ff}=
({32G\rho /3\pi})^{-1/2}$ reduces to $\sim8.08\times10^{6}/\rho^{1/2}$ yrs.
Finally, from (3) and (4) one can write $\tau_{coll}=\tau_{s}\Re$ 
where $\tau_{s}=\lambda/c_{s}\sim 1.07\times10^{5}\lambda(T_{4}/\mu)^{-1/2}$ 
yrs, $c_{s}$ is the sound speed, $\lambda$ is in pc,
and $\Re$ is the Mach number ($2V/c_{s}$) and is $\gg 1$. Eliminating the 
density using 
(7) the relevant time scale ratios become 
\begin{equation}
\tau_{cool}/\tau_{coll} < 
\tau_{cool}/\tau_{s} \sim  1.07\times10^{-5}\lambda(Z/Z_{\odot})^{-1}
(T_{4}/\mu)^{1/2} 
\end{equation}
and 
\begin{equation}
\tau_
{cool}/\tau_{ff}\sim 3.5\times10^{-5}\lambda(Z/Z_{\odot})^{-1}(T_{4}/\mu)^
{1/2}
\end{equation}
As will be 
shown in the next section, the observationally determined  value of $\lambda$ 
is such that both of these ratios are less 
than unity. Hence our assumption that the shock is isothermal is justified. We 
are also assuming that the temperature of the gas remains constant at the 
maximum of  
the cooling function near T$_{4}\sim1$ until the mass becomes Jeans unstable.
Further, one can expect fragmentation to occur when the gravitationally 
unstable cloud begins to collapse as long as $Z/Z_{\odot}$ is not much less 
than $\sim 0.01$. This is also the threshold metallicity below 
which globular clusters are not observed. 

Assuming that stars are eventually formed, another 
consideration is the efficiency $\eta$ with which gas is turned into stars. 
This is not well known at present so for the later discussion we 
assume this quantity to be $\eta \sim\Omega_{star}/\Omega_{baryon}\sim0.07$.
This ratio represents the global stellar mass density divided by the 
total baryon density. A further question then arises because the mass loss 
associated with star formation and evolution can lead to the disolution of any
cluster of stars formed. This problem has been investigated by Geyer \&
Burkert (2001) and more recently in their discussion of the globular cluster 
mass function by Parmentier \& Gilmore (2007). Unfortunately the observations 
which would allow an empirical estimate of the bound to unbound stellar ratio 
are not yet available so that more sophisticated modelling is required to 
answer this question in the context here. To proceed we shall 
consider $\eta$ to set an upper limit to the mass of a bound cluster of stars 
formed. 

A final point to note is that in this model the gas is concentrated by 
collision and the co-orbiting stars and dark matter will at least initially be 
unaffected by this process. Hence we do not expect the resulting globular 
clusters to contain a significant amount of dark matter.

\subsection{Relating the model to the Observations}

It is clear from the above discussion that the crucial quantity to determine 
is $\lambda$ while the principal variables are a characteristic length l, 
and a characteristic velocity V. Given that the available observational 
parameters of the dominant gaseous component of galaxies are the radius of the
HI disk, r$_{HI}$, the mass of the HI gas, M$_{HI}$, and the peak rotational 
velocity of the HI, V$_{rot}$, we begin by identifying l with r$_{HI}$ and V 
with V$_{rot}$. By identifying V with V$_{rot}$ we are considering mainly 
edge-on collisions. While the mass does not directly enter into the discussion 
above, 
correlations between it and the other variables lead to interesting results. 
The mass parameter actually used here is 1.3M$_{HI}$ in order to allow for the 
contribution of helium. Throughout the rest of this work, the symbol M$_{HI}$ 
will generally represent both the hydrogen $and$ helium gas content.

The data discussed here comes from two sources. The 
first for relatively bright galaxies are from Broeils and Rhee (1997) and the 
second for dwarf irregular galaxies are from Begum et al.(2008a,b). In order 
to 
intercompare both data sets a correction to both r and M was made for the small
difference in assumed Hubble constant. In addition because the HI radii 
were measured at a different limiting surface brightness, a correction of 
+0.178 in log r was applied to the Broeils and Rhee data. This was determined 
by jointly correlating the radii and the masses and finding the minimum 
dispersion. This correction is well within the uncertainties quoted by Begum 
et al. in their discussion of this problem. Fig. 2 shows the resulting relation
between 2 log r$_{HI}$ and log M$_{HI}$. The solid line has a slope of 1 while
a least squares fit for this slope gives 0.994$~\pm~0.012$ for all 137 
galaxies. The logarithmic intercept (log M-2log r) is 0.846$~\pm~0.132$. Hence
in agreement
with Begum et al. the HI surface density appears to be remarkably constant 
over nearly 5 orders of magnitude in mass. Note that this face-on surface 
density is not to be confused with $\Sigma_{o}$ above where it represents the 
column density.  

Equation (6) above implies that if $\lambda$ is a constant then a relation 
between 2 log V and log r should determine its value. Fig.3 shows such a 
plot. Possibly a more familiar way to look at this relation is to eliminate 
r by using the relationship established in Fig.2. Since r$\propto M^{1/2}$ 
equation (6) implies M$\propto V_{rot}^{4}$ i.e. an HI Tully-Fisher relation 
is predicted. This is shown in Fig.4. The solid line in both Figs 3 \& 4 has 
a slope of 1. Least squares fit to the data with V$_{rot} > 41.7~km~s^{-1}$ 
(i.e. above the dashed lines) give 0.978$~\pm~0.078$ in Fig.3 and 0.989$~\pm~ 
0.089$ in Fig. 4. Both regressions included 109 galaxies. The logarithmic 
intercepts are (log r-2 log V$_{rot}$)=
0.168$~\pm~0.250$ and (log M-4 log V)=1.18$~\pm~0.515$. Solving for $\lambda$ 
from each one obtains $\lambda=61.3~\pm~35.3(T_{4}/\mu)~pc$ and $\lambda=61.4
~\pm~36.3(T_{4}/\mu)~pc$ respectively. Although the uncertainties are large 
the numerical difference is small given the tight correlation seen in Fig. 2. 
Given this value of $\lambda$ the time scale arguments above are justified. It
should
be noted however that the condition for fragmentation will be violated for 
values 
of Z/Z$_{\odot}$ not much below 0.01. This provides a natural explanation for 
the observed abundance threshold of this order below which globular clusters 
are not observed a point also made by Gunn (1980).

Returning to equation (9) the Jean's mass becomes M$_{J}=3.7~\pm~2.1\times10^
{6}(T_{4}/\mu)^{2}~M_{\odot}$. Allowing for a gas to star formation efficiency
 of $\eta\sim0.07$ results in a 
predicted globular cluster mass upper limit of M$_{GC}\sim2.6~\pm~1.5\times10^
{5}(T_{4}/\mu)^{2}~M_{\odot}$.

Begum et al.~(2008a) have also noted that the dwarf irregular galaxies with 
the lowest rotational 
velocities (i.e. those below the dashed line in Figs 3 \& 4) lie below the 
extrapolated baryonic Tully-Fisher relation. If all of the galaxies are 
rotationally 
supported, then V$_{rot}^{2}=\alpha GM_{HI}/r_{HI}$ where $\alpha$ is the ratio
of the total mass enclosed within r$_{HI}$ to the gas mass. Given the 
relatively tight relation in Fig.2 the increased scatter in Figs 3 \& 4 can be 
understood as due to variations in this ratio. In order to account for the data
below the dashed lines, either the gas is not fully rotationally supported or 
there is a systematic shift in the value of $\alpha$ towards a lower value
inspite of the tight correlation between the HI radius and mass in fig. 2. 

Suppose that the galaxies below the dashed lines are indeed relatively gas 
rich (i.e. low $\alpha$). These objects could then be more representative of 
the universe as it was ~13 Gyr ago and be the
proto-galactic clumps from which the present galaxies were assembled. Fitting 
lines of slope unity through these 28 points in figs 3 and 4 then yields 
values of $\lambda$ of $310~\pm~272(T_{4}/\mu)~pc$ and $307~\pm~271(T_{4}/\mu)
~pc$ repectively 
i.e. a factor of $\sim$5 higher than from the more massive galaxies and hence 
a Jean's mass higher by the same amount. In order to recover the observed 
globular cluster masses one would need to invoke a star formation efficiency 
smaller by the same amount. In any event because the clusters are old and 
relatively metal poor, it will be argued below that these clusters were formed 
from the mergers of dwarf irregular galaxies whether those below the dashed 
lines in figs 3 and 4 or at the low mass end of those above.

At the beginning of this discussion we chose to identify the l and V of the 
model with r$_{HI}$ and V$_{rot}$ of the gas rich galaxies. The choice of 
V$_{rot}$ would seem to be natural since during the interaction between two 
counter rotating galaxies the undisturbed gas is rotationally supported.
The choice of r$_{HI}$ for l then leads to a `reasonable' mass for a globular 
cluster. A further consistency check comes from the model 
prediction that the column density of gas ($\Sigma_{o}=\rho_{o}$l) must be a 
constant 
and equal to $\Sigma_{o}=3.02\times10^{4}(T_{4}/\mu)/\lambda$. With the value 
of $\lambda$ derived above (assuming l=r$_{HI}$) $\Sigma_{o}=496~\pm~285
~M_{\odot}
pc^{-2}$. This in turn implies a value for the undisturbed particle density 
varying from $\sim7-0.2$ gas particles/cm$^{3}$ as the galaxy 
radius increases
from 2.5 to 100~kpc. For a Milky Way sized galaxy, (r$_{HI}\sim25$kpc), n$_{H}
\sim~0.6~\pm~0.4~cm^{-3}$ compared to an observed value of n$_{H}=
0.57~cm^{-3}$ (Dickey \& Lockman 1990). 

If structure in the universe forms hierarchically as is currently believed i.e.
low mass structures form first and merge to make larger objects, then we 
expect the oldest clusters to have been formed at the time when the low mass  
systems were merging. In fact genuine globular clusters (both blue and red) 
are found in many 
`surviving' dwarf irregular galaxies (e.g. van den Bergh 2006, Georgiev et 
al. 2008). Generally, as larger structures merge, we expect clusters of similar 
mass to form but have progressively younger ages and higher metallicities.
Today, 
given the right circumstances, these objects should still be forming and may 
resemble the super star clusters first observed by Schweizer (1987) in the 
Antennae galaxies. Conversely, the Galactic building blocks could not have 
been much more massive than the dwarf irregular galaxies before merging
or the red globular cluster backbone would have been younger and more metal 
rich than observed.
 
In the next section an outline of galaxy formation based on this picture is 
presented.

\section{Implications for Galaxy Evolution}

We now incorporate the above cluster formation model into an outline for 
galaxy evolution. The two separate globular cluster families (i.e. the old 
metal poor blue ones and the slightly younger metal rich red ones) are also 
found in different locations in a galaxy. Unlike the red clusters which are 
almost exclusively found in the inner regions of galaxy halos, the blue 
globular clusters can be found in both the inner and outer regions (e.g. Harris,
2001). In order to account for these 
differences we invoke an anisotropic collapse picture. We imagine that a 
spiral galaxy forms within a filament of proto-galactic clumps of gas and dark
matter. The filament is assumed to 
collapse first in a direction perpendicular to the filament length and then 
along its length. Generally the angular momentum content of a galaxy is 
distributed so we expect the low angular momentum building blocks to collide 
in the first collapse and in a major burst 
of star formation form the blue clusters and associated field stars.
While the previously formed stars and dark matter will form a more 
diffuse background, the gas left over after the star formation burst from 
which the blue globulars and associated field stars form will fall to the 
center and eventually be available to form the nucleus of the galaxy and 
contribute to the low angular momentum component of the bulge. The end result 
is a population of field stars formed prior to the mergers and their 
associated dark matter halos and post-merger blue globulars and 
stars formed as a result of the merging activity but with different kinematics
than the prior population. The higher 
angular momentum building blocks continue to become enriched and slowly spiral 
in towards the center at which time they begin colliding with each other and 
leave behind a diffuse background of stars, dark matter and any previously
formed blue globular clusters. The gas undergoes a second major burst of star 
formation and red globular cluster formation before the residual gas 
eventually falls to form 
the disk. A chemical evolutionary model based on this picture which 
was also able to account for the global star formation history was given in
H04. Here we compare the model to new and previously unavailable data based on
a large 
survey of stars with photometric metallicity determinations from the Sloan 
Digital Sky Survey (Ivezic et al. 2008).

Before discussing the observations we shall describe the chemical evolution 
model used to make the comparisons. It is similar to the model described in
H04. There the equations are modified slightly from a one zone mass loss model
in order to mimic a multi-zone model which works as follows. Suppose 
that in the one zone model the chemical evolution is interrupted before all 
of the gas has turned into stars as might occur during a merger for example. 
We are then left with a sharply truncated metallicity distribution of 
stars and a mass of gas with a single metallicity. Now suppose that 
these truncations occur over a finite range of metallicity dictated by a 
Gaussian distribution for example. 
Then we shall end 
up with a more smoothly truncated stellar distribution and finite spread in 
the metallicity of the left over gas which should be almost Gaussian shaped. 
The end result mimics the result of many zones having undergone truncations 
(mergers) over a range of metallicity. The model also allows for mass loss to 
occur at a rate proportional to the star formation rate by defining an 
effective yield.

Fig. 5 is taken from the paper of Ivezic et al.~(2008) (bottom right of their 
fig.
7). The points with error bars show the metallicity distribution of halo stars
with height above the plane $\mid Z\mid$=5-7 kpc. The blue curve shows the 
predicted 
metallicity distribution of the left over gas after the collisions have taken 
place. Because the model does not predict either the metallicity distribution 
or the mass of the stars formed during the mergers we have assumed that the 
post-merger stellar and cluster distribution is the same as that of the left 
over gas and used a gas to star formation efficiency of $\sim 0.07$ to 
estimate the mass of the stars and clusters. The solid black line (arbitrary 
normalization) shows the distribution of metallicity of the stars 
formed prior to the mergers. It was calculated using exactly the same chemical
evolution model given in H04. The model parameters are log(p$_{eff}/Z_{\odot})
=-0.3$, log($Z_{c}/Z_{\odot})=-1.50$, and $\sigma_{blue}=0.32$. Using the above
star formation efficiency, the ratio of the mass of 
stars formed during the mergers to those formed prior to the collisions is 
$\sim 0.8$. This number will also depend on the value 
assumed for p$_{eff}$. For illustrative purposes here and to minimize 
parameters we have kept the effective yield the same as the true yield (i.e. 
0.5Z$_{\odot}$). The stars formed prior to the merger are clearly more 
metal poor than the post-merger population and apparently do not show up in
the data in Fig. 5 although no attempt was made to fit more than one 
component to the data. These stars are expected to have different kinematic 
properties since they would not have been affected in the same way by the 
collisions which produced the burst of star formation and clusters. In fact, 
these stars may belong to the second blue halo population which is more metal 
poor and with different kinematics identified in a deeper survey by Carollo et
al.~(2007). 

Note that a prediction of this model is that generally the stars enclosed by 
the blue curve should have the same kinematics and spatial 
distribution as the blue 
(halo) globular clusters since they are assumed to have formed at the same 
time. Further these clusters should belong to the `young' halo group of 
Zinn (1993).

Fig. 6 from the top right of fig 7 of Ivezic et al. shows a similar plot but 
now with stars from $\mid Z\mid=1.5-2.0$ kpc. Here the red curve is the 
chemical model metallicity prediction assumed for stars and red clusters 
formed during the second set of collisions. Instead of using a Gaussian to 
determine the metallicity distribution of the stars/clusters formed during the 
collisions, we have used an extreme value distribution function 
which has the shape of an asymmetrical Gaussian. i.e. the function f of H04 is 
replaced by 
\begin{equation}
f=exp(-exp(-(log~Z_{c}-log~Z)/w))
\end{equation}
and the Gaussian function in equation (8) of H04 (df/dlogZ) is replaced by the
expression 
\begin{equation}
df/dlog~Z=exp((-(log~Z_{c}-log~Z)/w)-exp(-(log~Z_{c}-log~Z)1/w))/w
\end{equation}
The model parameters are log(p$_{eff}/Z_{\odot})
=-0.3$, log($Z_{c}/Z_{\odot})=-0.68$, and $w=0.16$. This function could have 
been used for the blue clusters as well but the fit was marginally better 
using the Gaussian for them given that only one component was being fit. 

The solid black 
line (arbitrary normalization) shows the metallicity distribution of all stars 
formed in the surviving population of building blocks prior to the second 
major merger phase. This group of stars (called the red spheroid in H04) will 
be more diffuse, have different kinematics and 
will possess a metal weak tail. The mass ratio between these two 
components (post to prior) for the parameters assumed above is $\sim 0.2$.

Note that any blue clusters that formed prior to the first major merger phase  
and which remained bound to its lower mass galaxy while avoiding the first 
mergers will also have the kinematics of the red spheroid population and 
hence would be members of the `old' halo component of Zinn (1993). The red 
clusters should have the same kinematics as the stars 
under the red curve of fig 6. The closest analog of a thick disk population 
in this model are these same stars. 

The main purpose of this discussion is to show that the phenomenological model 
outlined above is at least qualitatively consistent with the new data and with 
our picture of globular cluster formation. Further, the general outline for 
galaxy evolution described above may help explain additional globular cluster 
related questions. 
For example, the blue-red cluster phenomenom is also present in elliptical 
galaxies even though they are morphologically quite different from the gas 
rich systems. Could it be that ellipticals
formed similarly but at the intersection of more than one filament? The 
increased frequency of collisions would convert more gas to stars and lead to 
a higher specific frequency of globular clusters. In both cases variations in 
the distribution of angular momemtum content could explain the differences in 
the relative frequency of blue to red clusters. We leave these interesting 
questions to future work.

\section{Conclusions}

We have given a quantitative model for the formation of globular clusters 
based on the early work of Gunn \& McCrea. Clusters are assumed to form in 
merger induced collsions. We make use of recent HI data for dwarf irregular 
galaxies and earlier observational work to derive the key parameters of the 
model. We then combined this model with a simple chemical evolutionary model 
to predict the metallicity distributions of stars formed before and during 
these mergers. These predictions were compared with relatively recent survey 
data and shown to be consistent with both models. As additional data becomes 
available it will 
be possible to refine these ideas into a more sophisticated picture for galaxy
evolution.

\acknowledgements

The author is grateful to Zeljko Ivezic for promptly sending the data 
contained in Fig. 7 of their important paper. He also wishes to thank Brent 
Tully for suggesting references to the literature, Ray Carlberg for helpful 
comments and the referee for a constructive report.

\clearpage

\begin{figure}
\plotone{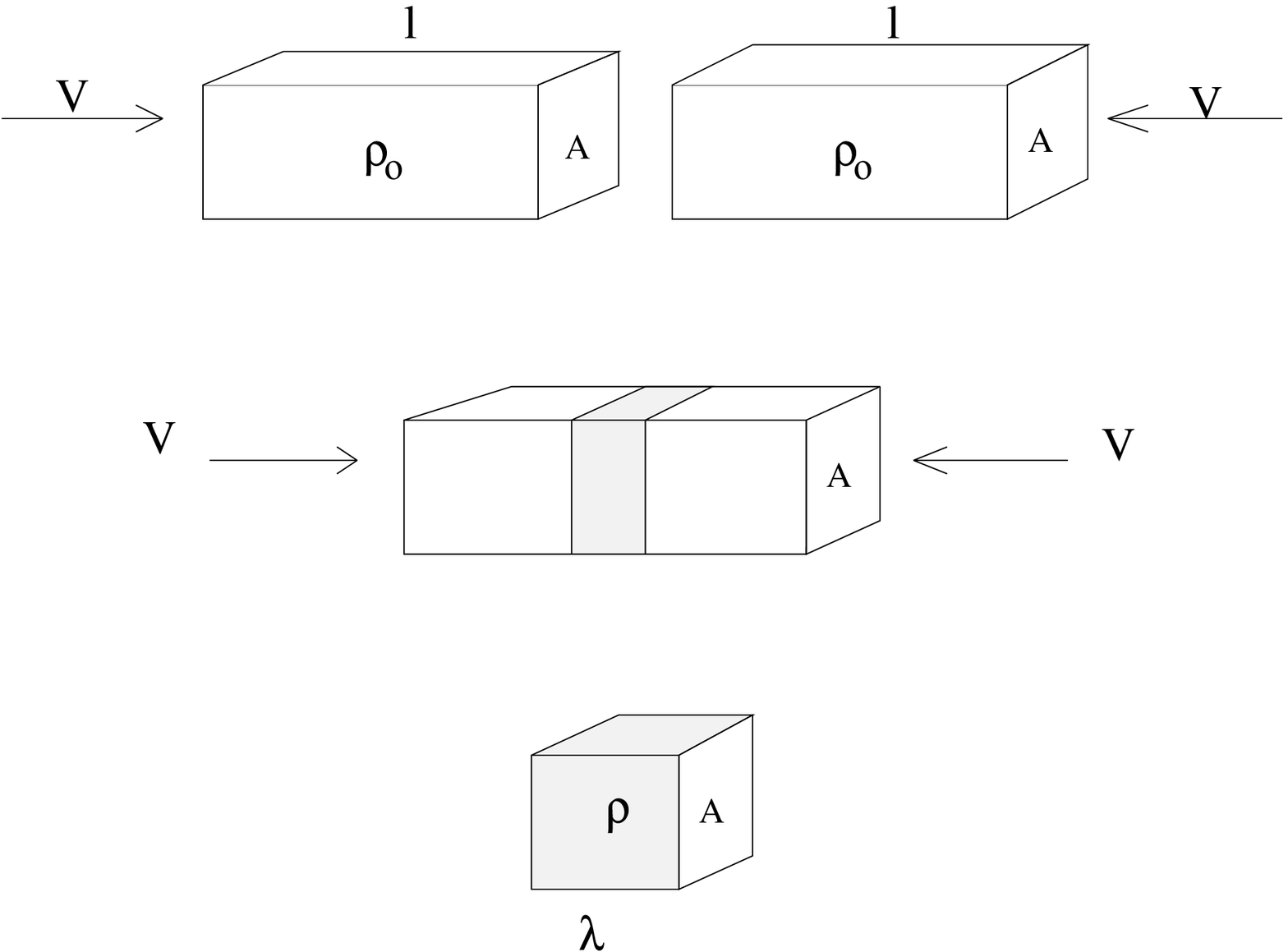}
\figcaption{Two gas columns approach one another with relative velocity 2V 
and collide reducing the column length from l to $\lambda$ and increasing 
the density from $\rho_{o}$ to $\rho$}
\end{figure}

\begin{figure}
\epsscale{.8}
\plotone{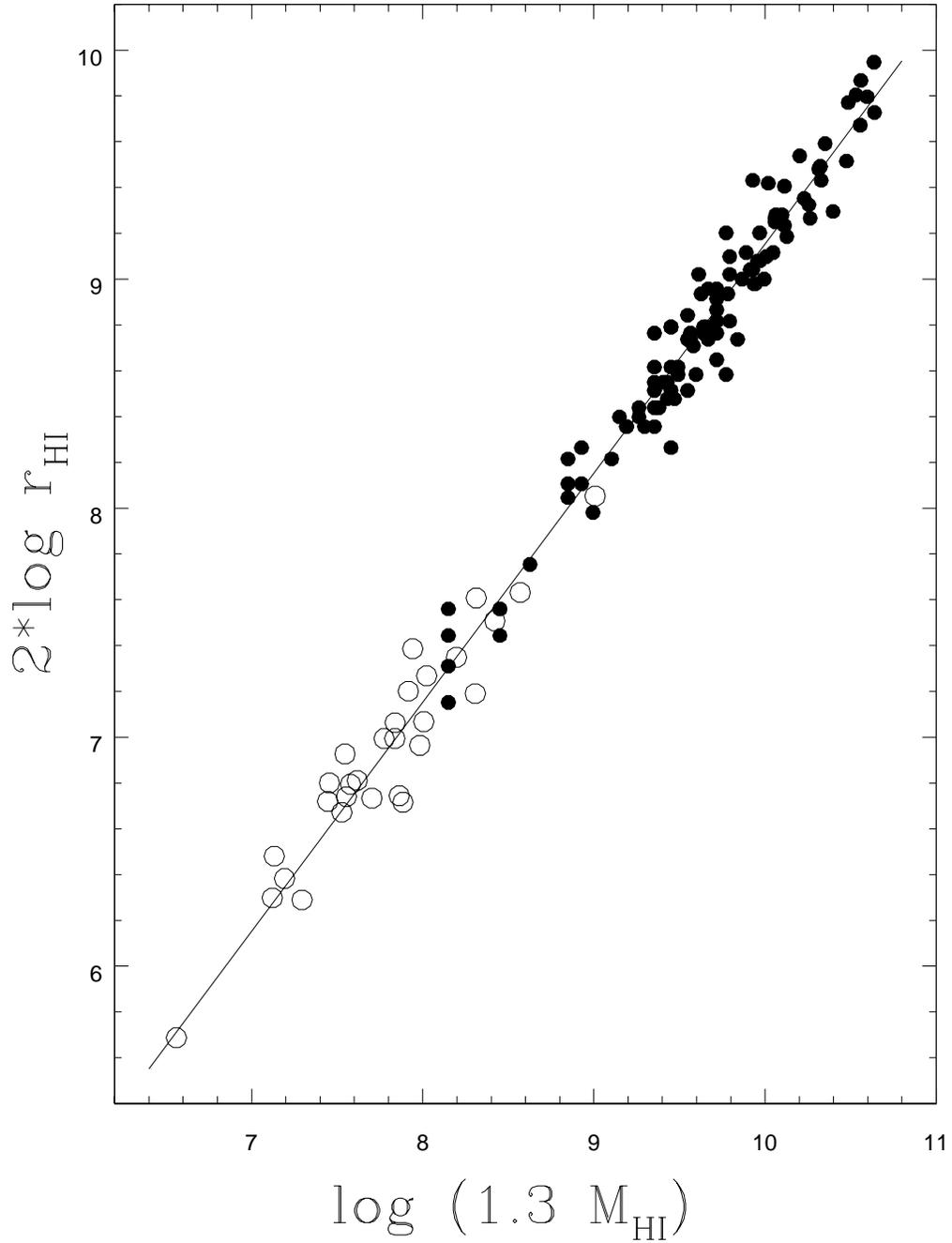}
\figcaption{Twice the log of the radial extent of the neutral hydrogen gas
r$_{HI}$ versus the log of 1.3$\times M_{HI}$. The solid line has a slope of 
unity. The solid points are from Broeils \& Rhee (1997) while the open circles
are from the work of Begum et al. (2008a,b). The units of r$_{HI}$ and $M_{HI}$
are pc and M$_{\odot}$ respectively.}
\end{figure}

\begin{figure}
\plotone{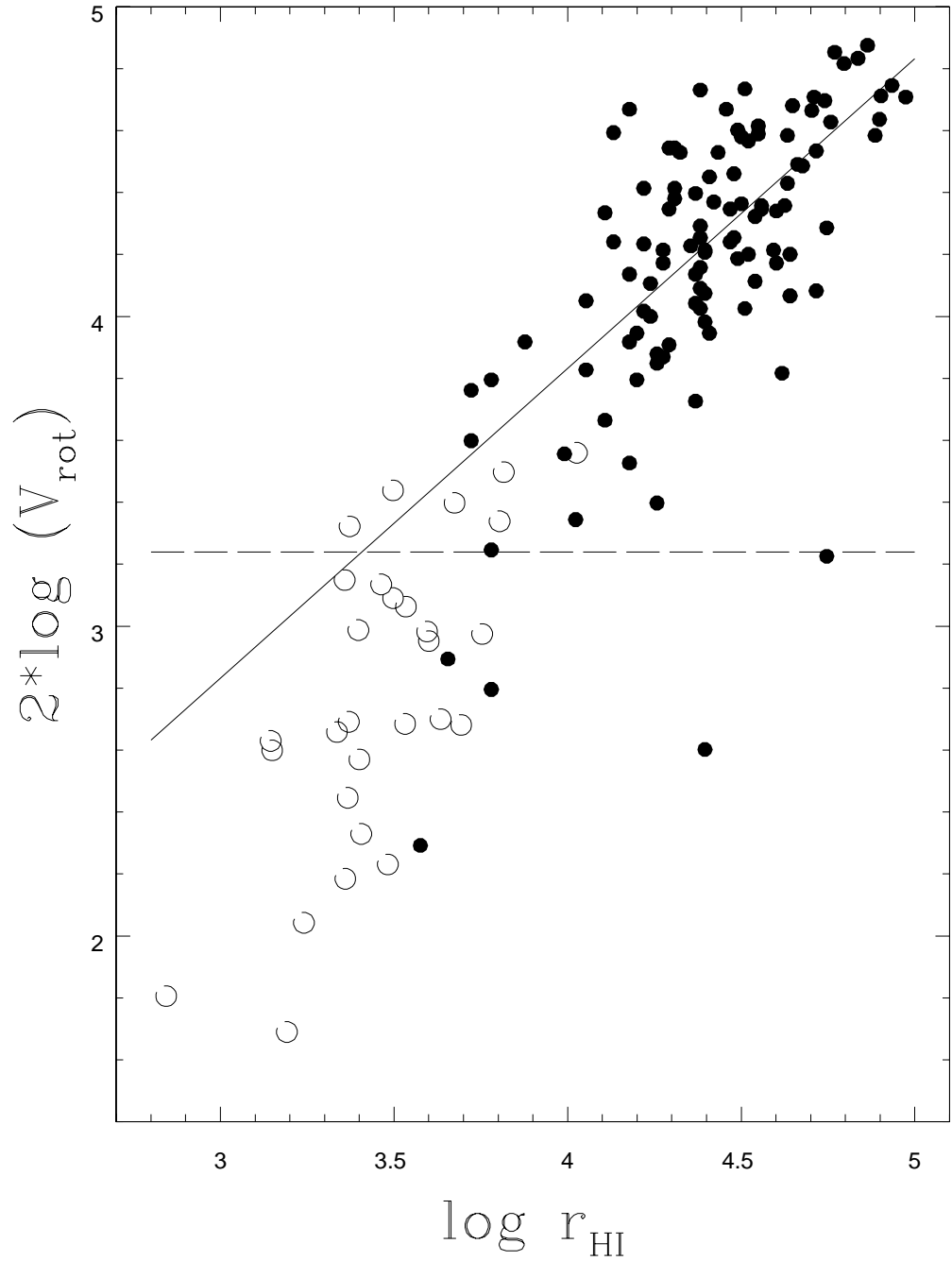}
\figcaption{Twice the log of the rotation velocity of the neutral hydrogen 
versus the log of its radial extent. The solid line has a slope of unity and 
the dashed line is the adopted lower limit of the velocity below which the 
predicted relation is assumed to break down. The symbols are as 
defined in fig. 2. The unit of rotational velocity is km/sec}
\end{figure}

\begin{figure}
\plotone{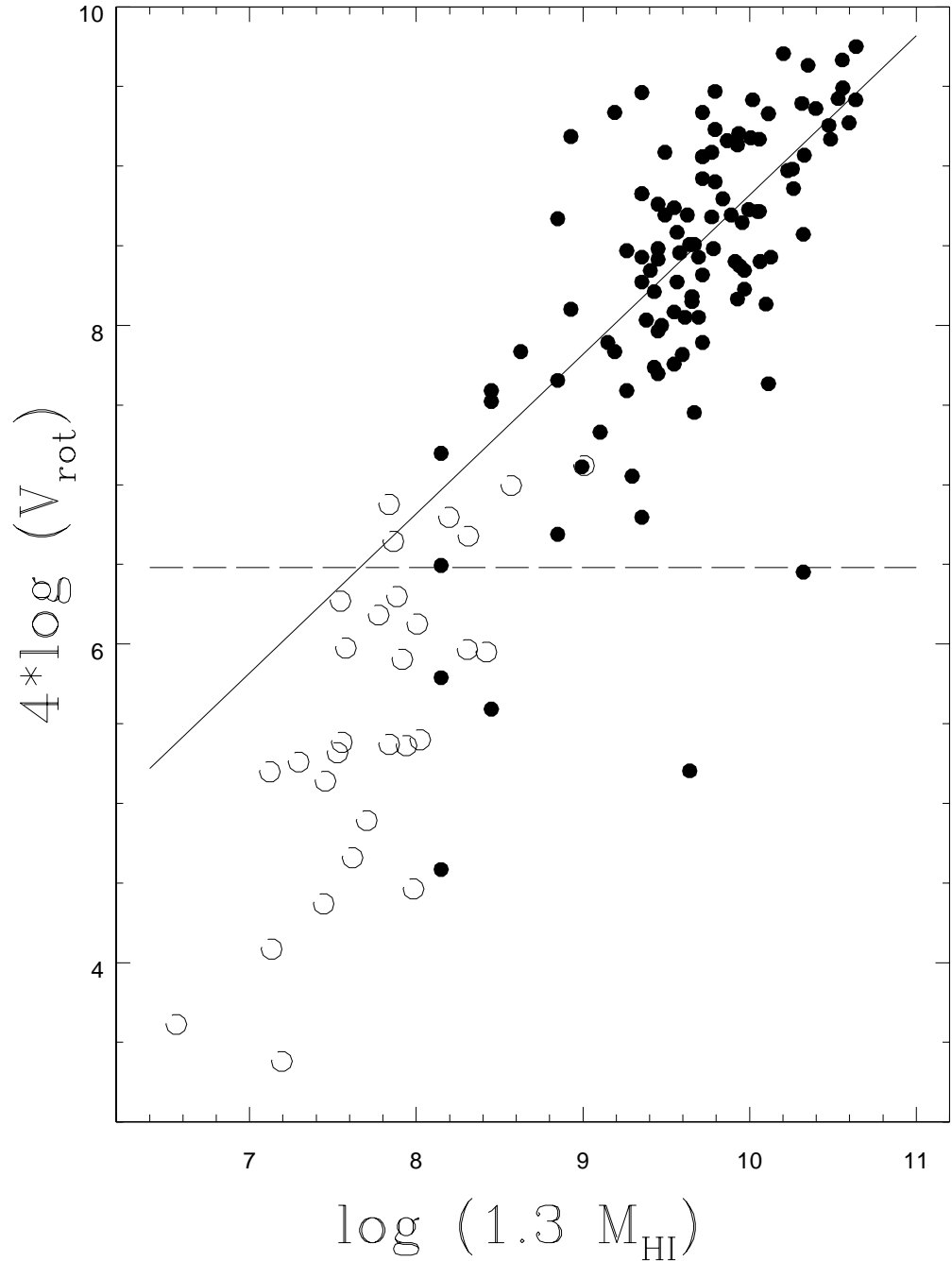}
\figcaption{Four times the log of the rotation velocity versus the log of the 
mass of gas. The solid line has a slope of unity and the dashed line is the 
adopted lower limit to the velocities of galaxies assumed to be obeying the HI 
Tully-Fisher relation. The symbols are as defined in fig.2.}
\end{figure}

\begin{figure}
\plotone{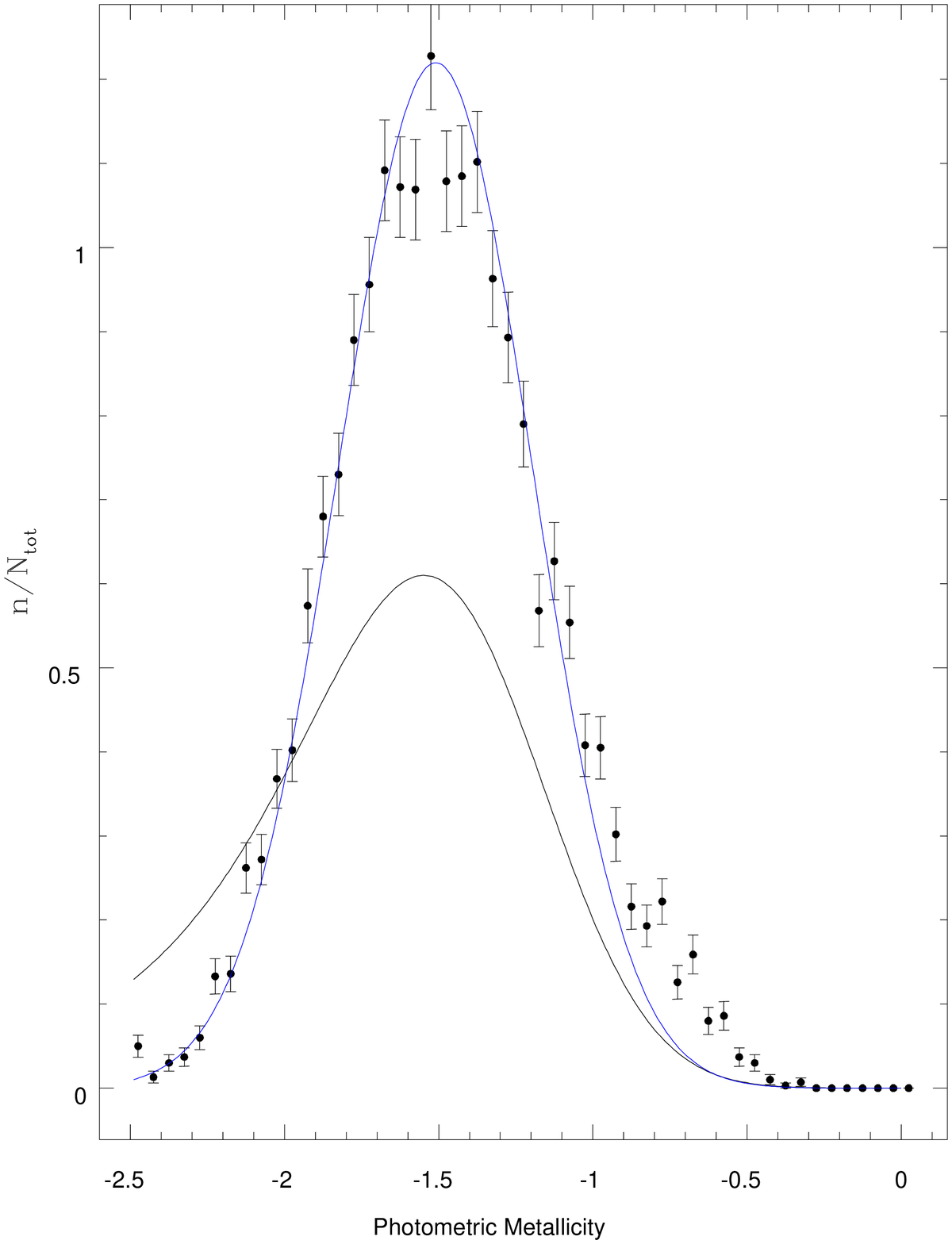}
\figcaption{The data points are from fig 7 (bottom right) of Ivezic et al. 
(2008) and show the metallicity distribution of stars within the distance 
interval above the plane of $\mid Z\mid=5-7$ kpc. The blue line is derived 
from the 
chemical evolution model under the assumption that the stars were formed in a 
major merger phase associated with the formation of the blue globular clusters.
The black line shows the predicted distribution of metallicities of the 
stars which formed prior to the merger phase. The 
normalization of this curve is arbitrary.}
\end{figure}

\begin{figure}
\plotone{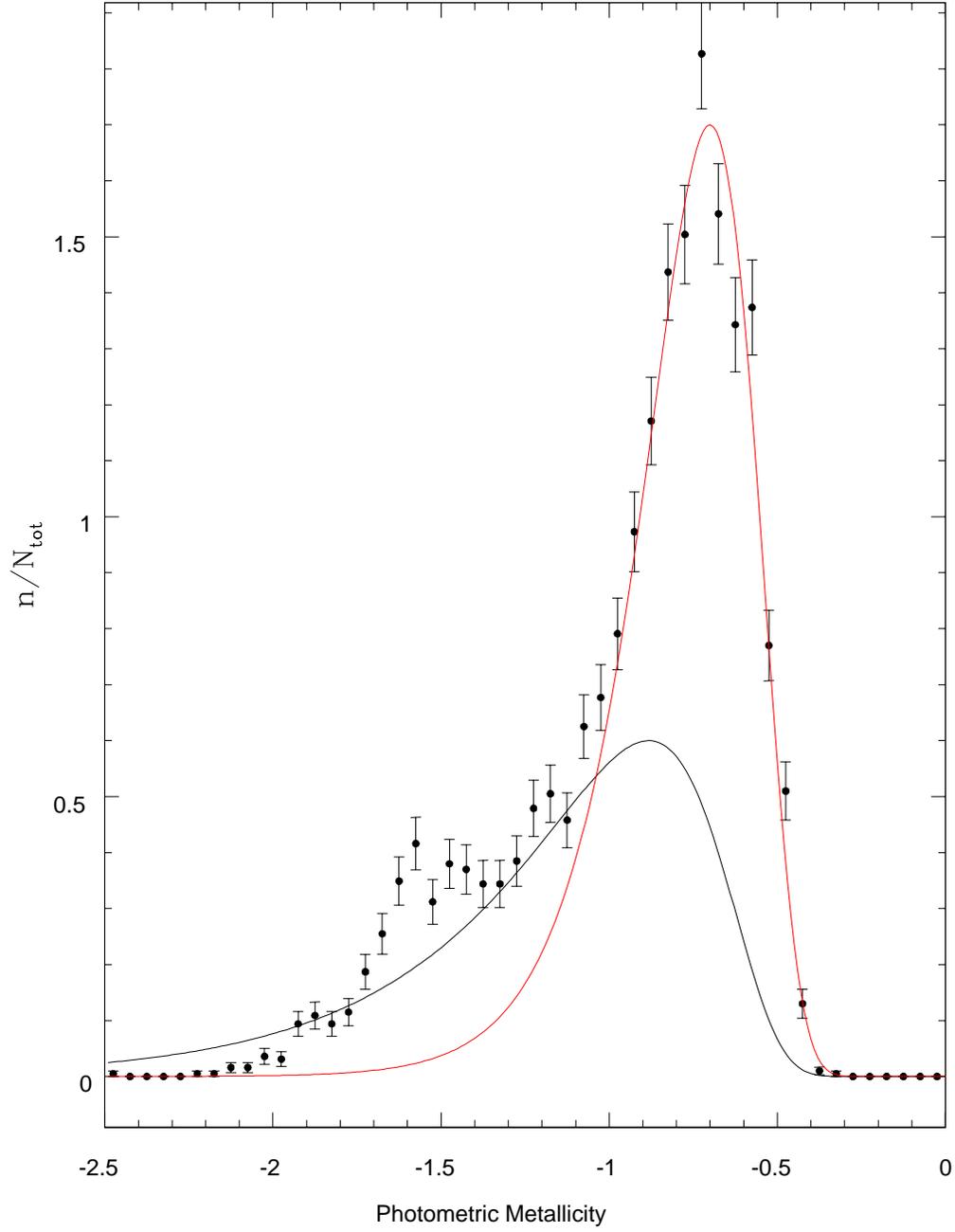}
\figcaption{The data points are from fig 7 (top right) of Ivezic et al. (2008) 
and show the metallicity distribution of stars within the distance interval 
above the plane of $\mid Z \mid=1.5-2.0$ kpc. The red line is derived from the 
chemical evolution model under the assumption that the stars were formed in a 
second major merger phase associated with the formation of the red globular 
clusters. As above 
the black line with arbitrary normalization shows the predicted distribution of
metallicities of stars formed prior to the mergers.}
\end{figure}

\end{document}